\documentclass[useAMS,usenatbib]{mn2e}
\usepackage{rotating}
\usepackage{graphics}
\usepackage{graphicx, subfigure}
\usepackage{tabularx}
\usepackage{amssymb}
\usepackage{setspace}
\usepackage{txfonts}
\usepackage{natbib}
\usepackage{longtable}
\usepackage{epstopdf} 
\usepackage{epsfig}

\title[UCD spectra with OSIRIS/GTC]{Low-resolution optical spectra of ultracool dwarfs with OSIRIS/GTC}
\author[Metodieva, Y. et al.]{Metodieva, Y.$^{1,2}$\thanks{Based on observations made with the Gran Telescopio Canarias (GTC), instaled in the Spanish Observatorio del Roque de los Muchachos of the Instituto de Astrof\'{\i}sica de Canarias, in the island of La Palma}\thanks{e-mail: ytm@arm.ac.uk}, Antonova, A.$^{1}$, Golev, V.$^{1}$, Dimitrov, D.$^{3}$, Garc\'{\i}a-\'Alvarez, D.$^{4,5,6}$ 
\newauthor and Doyle, J. G.$^{2}$ \\
$^{1}$Department of Astronomy, St. Kliment Ohridski University of Sofia, 5 James Bourchier Blvd., 1164 Sofia, Bulgaria\\
$^{2}$Armagh Observatory, College Hill, Armagh BT61 9DG, UK\\
$^{3}$Institute of Astronomy and NAO, Bulgarian Academy of Science, 72 Tsarigradsko Chaussee Blvd., 1784 Sofia, Bulgaria\\
$^{4}$Instituto de Astrof\'{\i}sica de Canarias, Avenida V\'{\i}a L\'actea, 38205 La Laguna, Tenerife, Spain\\
$^{5}$Departamento de Astrof\'{\i}sica, Universidad de La Laguna, 38205 La Laguna, Tenerife, Spain\\
$^{6}$Grantecan S.\,A., Centro de Astrof\'{\i}sica de La Palma, Cuesta de San Jos\'e, 38712 Bre\~na Baja, La Palma, Spain\\
}

\begin{document}

\date{Accepted 2014 November 6. Received 2014 November 6; in original from 2014 June 13}

\pagerange{\pageref{firstpage}--\pageref{lastpage}} \pubyear{2014}

\maketitle

\label{firstpage}

\begin{abstract}
We present the results of low-resolution optical spectroscopy with OSIRIS/GTC (Optical System for Imaging and Low Resolution Integrated Spectroscopy / Gran Telescopio Canarias) for a sample of ultracool dwarfs. For a subsample of seven 
objects, based on 2MASS NIR photometric colours, a 'photometric' spectral type is determined and compared to the results of the optical spectroscopy. For the stars, showing H$\alpha$ line in emission, equivalent widths were measured, and the ratio of H$\alpha$ to bolometric luminosity were calculated. We find that two dwarfs show the presence of magnetic activity over long periods, LP 326-21 -- quasi-constant-like, and 2MASS J17071830+6439331 -- variable. 
\end{abstract}

\begin{keywords}
 stars: activity -- brown dwarfs -- stars: chromospheres -- stars: low-mass.
\end{keywords}

\section{Introduction}

Late-type dwarfs are the most populous objects in the Galaxy, comprising roughly 70 per cent of all stars in the Solar neighbourhood and contributing around half of the Galactic stellar mass \citep{Henry97}. Despite their low mass and effective temperature, some of these dwarfs reveal strong magnetic activity, orders of magnitude stronger than the Sun's. In the past few decades using surveys like 2 Micron Sky Survey (2MASS) and Sloan Digital Sky Survey (SDSS) a large number of such objects were discovered. To date, the determination of their physical properties is of great significance for the characterization of the population of low mass stars in the Galaxy and especially in the Solar neighbourhood. The most common characteristic of magnetic activity in cool dwarf stars is seen in chromospheric lines such as Ca~{\sc ii} H and K and H$\alpha$ or as coronal X-ray emission. Among the M dwarfs there is a well-known empirical distinction between the two classes (dM and dMe) based on the optical criterion: dMe stars have H$\alpha$ in emission, while dM do not. \

\citet{CramMullan79} showed that the source of emission arises from a chromosphere of increasing density. This higher density causes the H$\alpha$ 
line to pass from radiative (H$\alpha$ in absorption) to electron collisional dominated (H$\alpha$ in emission). Furthermore, \citet{CramGiampapa} 
noted that H$\alpha$ in absorption in late-type dwarfs also requires the presence of a non-radiatively heated chromosphere since the photospheric radiation field is not strong enough to populate the \textit{n} = 2 and higher levels of the hydrogen atom. Work by \citet{Byrne93} draw attention to those M 
dwarfs with no detectable H$\alpha$ emission. Such stars, instead of being inactive, may have the H$\alpha$ absorption line filled-in.   \

For the dwarfs showing H$\alpha$ in emission, the activity levels are seen to increase towards later types due to longer spin-down time-scales, reaching a peak at spectral type $\sim$ M7 \citep{west04}. Beyond spectral type M7, there is a drop-off in H$\alpha$ and X-ray activity, with a rapid decline seen for L and T dwarfs \citep{gizis00, mohanty03, fleming03, schmidt07}. This suggests a decrease in the generation and dissipation of magnetic fields for late-M and L dwarfs. One reason could be the increasing atmospheric neutrality in ultracool dwarfs that decouples the field from atmospheric motions, and thus hamper their dissipation even if strong fields are generated. This decoupling can cause suppression of chromospheric and coronal heating, resulting in a decrease in emission in the H$\alpha$ line and X-rays, despite the presence of magnetic fields \citep{mohanty02}. Indeed, magnetic field strengths of kG have been measured for some late-M and L dwarfs indicating that the drop in H$\alpha$ emission seen for ultracool dwarfs (dwarfs with spectral type $\geq$ M7) must be due to atmospheric neutrality rather than a fall in magnetic flux production \citep{reiners07, hallinan08}. This is supported also by the fact that both flaring and quiescent H$\alpha$ emission is still detectable from a number of dwarfs of spectral types L and T albeit with luminosities much lower than those of earlier M type dwarfs \citep{reiners10,burgasser03,liebert03}. \
\begin{table}
\centering
\caption{Spectral type--infrared colours relation for late-type M dwarfs, as given by the polynomial fit based on photometry--spectral type calibration using 367 known dwarfs in the spectral type range M6--L0, taken from the DwarfArchives.org.} \
\label{colours}
\begin{tabular}{c c c c}
\hline

SpT	&	$J-H$	&	$H-K$	&	$J-K$ \\

\hline

M6.5	&	0.625	&	0.378	&	1.003 \\
M7.0	&	0.649	&	0.399	&	1.048 \\
M7.5	&	0.659	&	0.416	&	1.075 \\
M8.0	&	0.676	&	0.440	&	1.116 \\
M8.5	&	0.696	&	0.457	&	1.153 \\
M9.0	&	0.719	&	0.475	&	1.194 \\
M9.5	&	0.742	&	0.493	&	1.235 \\
L0.0	&	0.765	&	0.512	&	1.277 \\

\hline
\end{tabular}
\end{table}

\begin{table*}
\begin{minipage}{180mm}
\caption{The full sample of dwarfs, selected for observation, their coordinates, spectral types and distances: 1-- \citet[][and references therein]{tony13}; ~2-- \citet{allen07}; ~3-- \citet{faherty09}; 4-- \citet[][and references therein]{mclean12}; ~5-- \citet{reid08}; ~6-- \citet{sion09} . }
\centering
\vspace*{0.3cm}
\label{objects}
\begin{tabular}{l c l l l l l l l l}
\hline

2MASS designation	&	Other name	&	RA		&	Dec		&	Sp. T.	&	Observation	&	Grism	&	Exposure	&	Distance	&	Ref. \\
			&			&	(J2000)		&	(J2000)		&	/lit./	&	date 		&		&	(s)		&	(pc)		&		   \\
\hline

J13540876+0846083 	&		--		&	13 54 08.77	&	+08 46 08.4	&	M7.0	&	2013-04-18	&	R1000B  &	2 x 600s      	&       17	      	&       1       \\
J14280419+1356137 	&	LHS 2919		&	14 28 04.20	&	+13 56 13.7	&	M8.0	&	2013-04-17	&	R1000B  &	2 x 400s      	&       10	      	&       1, 4    \\
J14441717+3002145 	&	LP 326-21	&	14 44 17.18	&	+30 02 14.5	&	M8.0	&	2012-12-22	&	R1000B  &	3 x 200s      	&       12.7	      	&       1, 2, 4 \\
J15164073+3910486 	&	LP 222-65	&	15 16 40.73	&	+39 10 48.7	&	M6.5	&	2012-09-19	&	R1000B  &	1 x 180s      	&       --	      	&       4       \\
J16063390+4054216 	&	LHS 3154		&	16 06 33.90	&	+40 54 21.6	&	M6.5	&	2013-01-11	&	R1000B  &	2 x 400s      	&       --	      	&       4       \\
J16241436+0029158 	&				&	16 24 14.37	&	+00 29 15.8	&	T6		&	2012-09-08	&	R1000B  &	2 x 400s     	&       11.0 $\pm$ 0.1  &       1, 3    \\
J16342164+5710082 	&	GJ 630.1B	&	16 34 21.64	&	+57 10 08.3	&	DQ8+dM4	&	2012-09-17	&	R1000B  &	2 x 400s      	&       14.5	      	&       4, 6    \\
J16463154+3434554 	&	LHS 3241		&	16 46 31.55	&	+34 34 55.5	&	M6.5	&	2012-09-13	&	R1000B  &	2 x 400s      	&       --	      	&       4       \\
J17071830+6439331 	&				&	17 07 18.31	&	+64 39 33.1	&	M9.0	&	2012-09-06	&    	R1000B  &	2 x 700s      	&       16.4 $\pm$ 1.1	&       1, 5    \\
	--				&		--		&		--	&		--			&	--		&	2013-04-17	&	R1000B	&	2 x 700s      	&		--	  &       1, 6    \\
J17462953+2903386 	&		--		&	17 46 29.53	&	+29 03 38.6	&	--		&	2013-04-25	&	R1000R  &	2 x 400s      	&       --	      	&     --         \\
J17465934+4502473 	&		--		&	17 46 59.34	&	+45 02 47.3	&	--		&	2013-04-25	&	R1000R  &	2 x 400s      	&       --	      	&       --       \\
J17502484-0016151~~	&		--		&	17 50 24.84	&	--00 16 15.1	&	L5.5		&	2013-02-08	&	R1000B  &	2 x 800s      	&       9 $\pm$ 1       &       1, 3    \\
J18573008+5015011 	&		--		&	18 57 30.08	&	+50 15 01.1	&	--		&	2013-04-25	&	R1000R  &	2 x 400s      	&       --	      	&     --         \\
J20015863+6427486 	&		--		&	20 01 58.63	&	+64 27 48.6	&	--		&	2013-04-26	&	R1000R  &	2 x 400s      	&       --	     	&       --      \\
J21512797+3547206 	&		--		&	21 51 27.97	&	+35 47 20.6	&	--		&	2013-05-31	&	R1000R  &	2 x 400s      	&	--	      	&        --      \\
J22594403+8013189 	&		--		&	22 59 44.03	&	+80 13 18.9	&	--		&	2013-05-31	&	R1000R  &	2 x 400s      	&       --	      	&         --    \\
J23333174+7456179 	&		--		&	23 33 31.74	&	+74 56 17.9	&	--		&	2013-05-31	&	R1000R  &	2 x 400s      	&       --	      	&          --    \\
	
\hline
\end{tabular}
\end{minipage}
\end{table*}

On the other hand, despite the observed sharp drop in H$\alpha$ and X-ray luminosities, the radio luminosities remain unchanged in the ultracool dwarf regime. What is more, some of the radio detected dwarfs exhibit rotational modulation of the signal and/or highly circularly polarized pulses indicative of a coherent mechanism -- the electron cyclotron maser (ECM) \citep{tony08, hallinan08, berger09, Doyle10, mclean11, Williams13} . For two of these dwarfs, rotational modulation was also found in H$\alpha$ \citep{berger08, berger09}, suggesting a common source, most likely an active spot on the surface \citep{Yu11, Alex12}. Another set of ultracool dwarfs exhibit a high degree of long-term variability, with emission levels dropping below the detection limit in some instances \citep{tony07, osten09, mclean12}.\

Late-type dwarfs are generally termed as magnetically active (or inactive) by the equivalent width of the H$\alpha$ emission line -- EW(H$\alpha$) \citep{walkow04,west05}; however, note that the earlier comments concerning a filled-in H$\alpha$ line.  Because the measurements of the EW(H$\alpha$) depend on the flux in the respective local continuum, EW(H$\alpha$) cannot be a measure for comparison of the strength of magnetic activity between stars of different spectral types. Instead the activity strength is measured by the ratio of the luminosity in H$\alpha$ to the bolometric luminosity (L$_{H\alpha}$/L$_{bol}$).\

Studies of the magnetic activity impose important limitations for the inner structure, the relations between generation of the magnetic field and rotation, atmospheric models and ages of late-type dwarfs. Previous work shows that the activity strength, measured by the ratio L$_{H\alpha}$/L$_{bol}$, is nearly constant (with large scatter) through the M0--M6 range \citep{hawley96} and declines at later types \citep{burgasser03, cruz+reid02}. \citet{west04, west08} and \citet{bochanski07} use data from SDSS to determine the properties (again via H$\alpha$) of the magnetically active dwarfs. \

Detection of H$\alpha$ emission can provide further information on the nature of their activity and the activity among ultracool dwarfs in general. Obtaining H$\alpha$ luminosities or upper limits and correlating them with available radio and X-ray luminosities and \textit{v}$\sin$\textit{i} measurements will help to better characterize the overall picture of activity in the ultracool dwarf regime (beyond the scope of this paper).\

\begin{figure}
\begin{center}
\epsfig{file=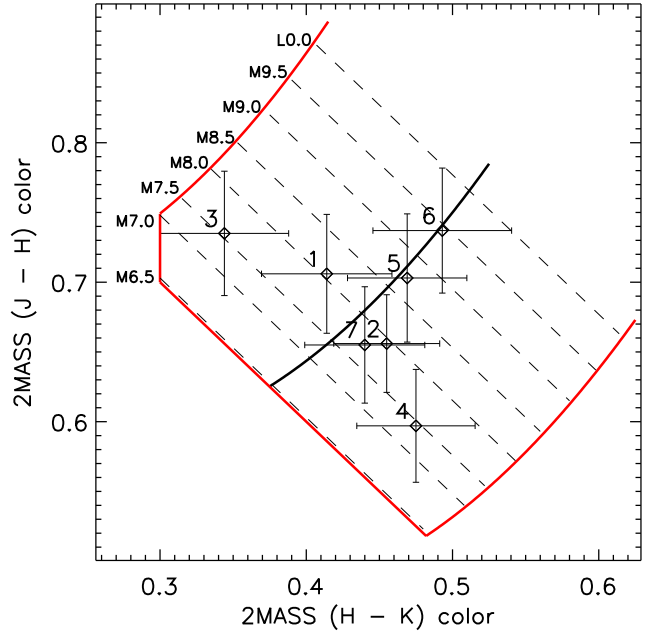, width=8.5cm}
\caption[]{Colour--colour diagram for the late-M dwarf candidates. The thick black line is the polynomial fit to the data for known dwarfs; red curves denote the 3$\sigma$ deviations from the fit; the straight red lines represent the requirements $(H-K) \geq$ 0.3, and $(J-K) \geq$ 1.0. The dashed black lines show the resultant spectral classes as listed in Table \ref{colours}. The candidates are: 1 -- 2MASS J1746+29, 2 -- 2MASS J1746+45, 3 -- 2MASS J1857+50, 4 -- 2MASS J2001+64, 5 -- 2MASS J2151+35, 6 -- 2MASS J2259+80, 7 -- 2MASS J2333+74.} \
\label{phot_types} 
\end{center}
\vspace{-0.5cm}
\end{figure}


\section{Target selection}
Here we present the results from spectral observations of 17 M, L and T dwarfs made with the Gran Telescopio Canarias, La Palma, Spain. These dwarfs are a subset of a much larger sample of UCDs including objects studied for other activity characteristics (such as radio and X-ray emission \citep[e.g.][]{tony13}) and several thousand candidates for late-M, L and T dwarfs, based on infrared colours \citep[e.g.][]{dinko11}. The selection of these particular objects is due to the constrains, imposed by the observation program. Observations were performed in service mode within the GTC `filler' programme GTC30-12B and GTC3-13A. The aim of this filler programme was to obtain, within the GTC nightly operation schedule, high-quality spectra of {\em variable} and/or sources with no spectral type determination that are relatively bright for a 10\,m-class telescope (with magnitudes of up to $V \sim$ 19\,mag) in poor weather conditions, such as dust presence, sky brightness, dense cirrus coverage, or poor seeing (of over 1.4\,arcsec). \

\subsection{Late-M candidates}
In this group we have seven objects, candidates for late-M dwarfs. The $J, H$ and $K$ colours are taken from the 2MASS catalogue, while $I$ is from the GSC 2.3 catalogue. The 2MASS photometry--spectral type calibration is based on 367 dwarfs in the spectral type range M6--L0, taken from the DwarfArchives.org \footnote{http://spider.ipac.caltech.edu/staff/davy/ARCHIVE/index.shtml}. For all these stars, there are both spectra and photometry available, from which an empirical relation between photometric colours and spectral type can be derived. The result of the polynomial fit to the known dwarfs is shown in Fig. \ref{phot_types}, while the NIR indices--spectral type relations are listed in Table \ref{colours}. \

The selection criteria, based on optical and NIR colour indices are:  $(J-K) \geq$ 1.0 mag, $(I-K) \geq$ 3.0 mag, 0.52 $\leq (J-H) \leq$ 0.9 and 0.30 $\leq (H-K) \leq$ 0.60. These requirements also avoid contamination by giants or galaxies. We've also constrained \textit{J} magnitudes to the range 10 $\leq$ J $\leq$ 14, since dwarfs with J $<$ 9 are likely to have already been studied by \citet{lepine13}, and dwarfs with J $>$ 14 will be to faint to satisfy the observation programme requirements. \

Another criteria that distinguishes giants from dwarfs is the proper motion, so only objects with proper motions greater than  0.030 $\arcsec$ year$^{-1}$ were considered. Proper motions are determined using two epochs of observations -- from the 2MASS catalogue and the STScI Digitized Sky Survey (POSSI Red plates). Based on the above calibration, a photometric spectral type is calculated for each object. In addition, we selected seven dwarfs which had a spectral classification but no H$\alpha$ measurements. \

\subsection{Looking for activity}

We have selected 10 objects out of $\sim$ 70 dwarfs, observed in the radio and/or X-ray domains and with spectral classification, but without  H$\alpha$ measurements \citep[][and references therein]{tony13}. The goal is to gain additional information about these objects and their overall magnetic activity. \

\section{Instruments, observations and data reduction} \label{observations}\

We carried out low-resolution spectroscopy with the Optical System for Imaging and Low Resolution Integrated Spectroscopy (OSIRIS) tunable imager and spectrograph \citep{Cepa03, Cepa10} at the 10.4\,m Gran Telescopio Canarias (GTC), located at the Observatorio Roque de los Muchachos in La Palma, Canary Islands, Spain. The heart of OSIRIS is a mosaic of two 4k\,$\times$\,2k e2v CCD44--82 detectors that gives an unvignetted field of view of 7.8\,$\times$\,7.8\,arcmin$^{2}$ with a plate scale of 0.127\,arcsec\,pixel$^{-1}$. However, to increase the signal-to-noise ratio (S/N) of our observations, we chose the standard operation mode of the instrument, which is a 2\,$\times$\,2-binning mode with a readout speed of 100\,kHz. All spectra were obtained with the OSIRIS R1000B (blue) and R1000R (red) grisms. The blue grism has a wavelength range of 3630 \AA ~ - 7500 \AA, centred on 5510 \AA, with a resolution of 1018 ($\sim$ 2.12 \AA\,pixel$^{-1}$) and maximum quantum efficiency of 65\%. The red grism has a wavelength range of 5100 \AA ~ - 10000 \AA, centred on 7510 \AA, with a resolution of 1122 ($\sim$ 2.62 \AA\,pixel$^{-1}$) and maximum quantum efficiency of 65 per cent.
We used the 1.23\,arcsec-width slit, oriented at the parallactic angle to minimize losses due to atmospheric dispersion. The resulting resolution, measured on arc lines, was R $\sim$ 700 in the approximate 3500--8000\,{\AA} spectral range. Observations were performed in service mode within the GTC `filler' programme GTC55--12A on different nights in the two observational seasons 2012B and 2013A. A detailed observational log is presented in Table \ref{objects}.\

 Data reduction for all spectra was done using the software package \textsc{idl} (Interactive Data Language) and its astronomical libraries \citep{landsman95}. The code was written especially for this work and thus a package for reducing OSIRIS spectra was constructed. The data reduction followed standard procedures, e.g. bias, flat-fielding, cosmic ray cleaning, wavelength calibration and flux calibration made using standard stars with well-known spectral energy distributions. \\

\section{Spectral classification and chromospheric activity}

\subsection{Spectral classification}

All of the objects are spectrally classified (Table~\ref{results}) using the software package \textsc{hammer} (http://www.astro.washington. edu/users/slh/hammer) developed by \citet{covey07}. This is an \textsc{idl} based package originally developed for use on late-type SDSS spectra, but has been modified for a number of other objects and works for all spectral classes from O5 to L8. Theoretically, the accuracy with which \textsc{hammer} works is about two subclasses \citep{covey07}, but after automatic and visual classifications made for this work, it was found that the visual spectral type is the same as the one from the \textsc{hammer} interactive mode ($\pm$~one subclass). The spectral types for the late-M dwarf candidates, assigned with \textsc{hammer}, are within 0.5 subclass of the ones, based on photometric colours, with the exception of 2MASS J2MASS J2259+80, for which the deviation is one subclass (see Fig. \ref{phot_types}).

\subsection{Chromospheric activity}

\begin{figure*}
\centering
$
\begin{array}{cc}
\includegraphics[width=5.5cm, angle=-90]{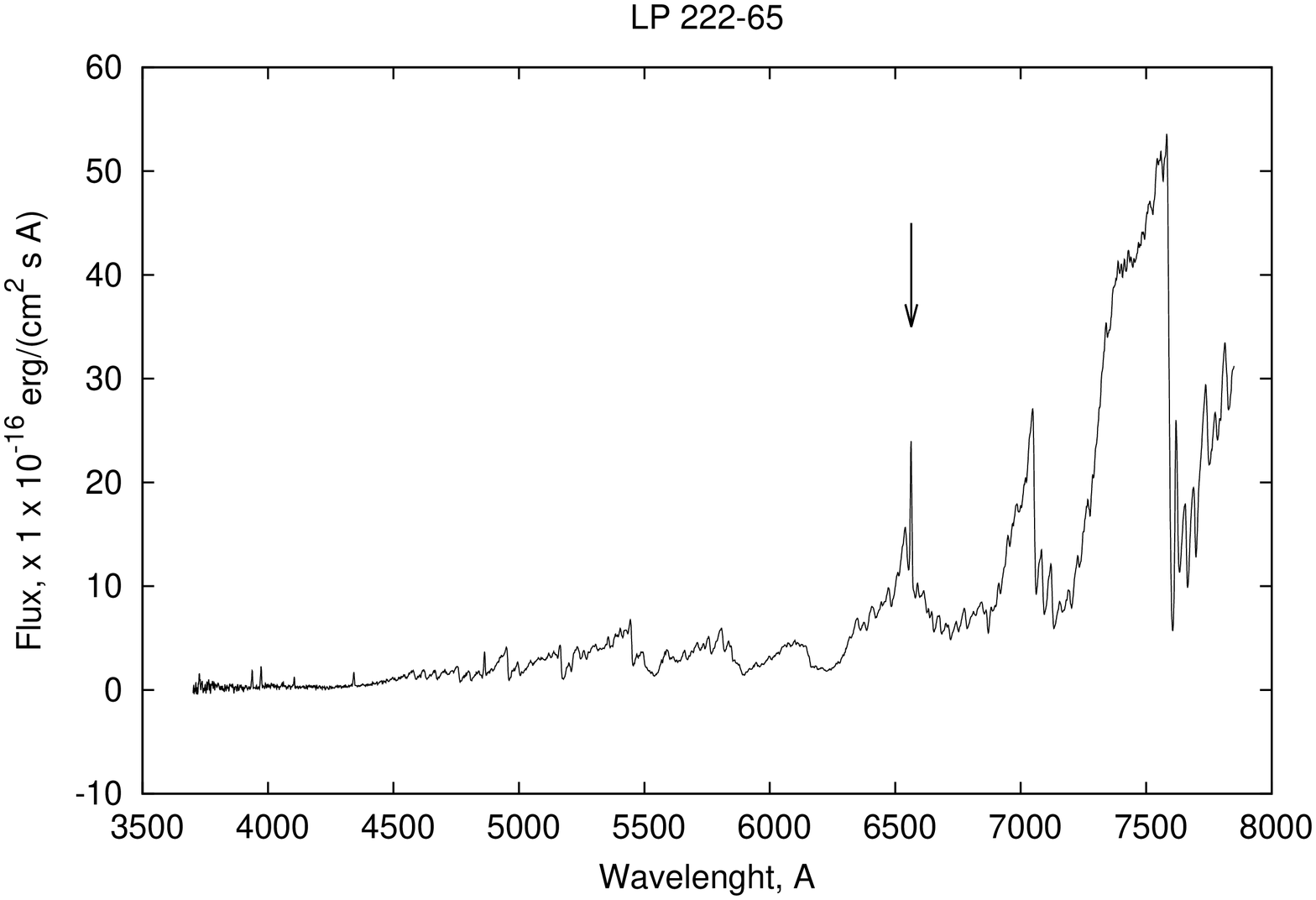} &
\includegraphics[width=5.5cm, angle=-90]{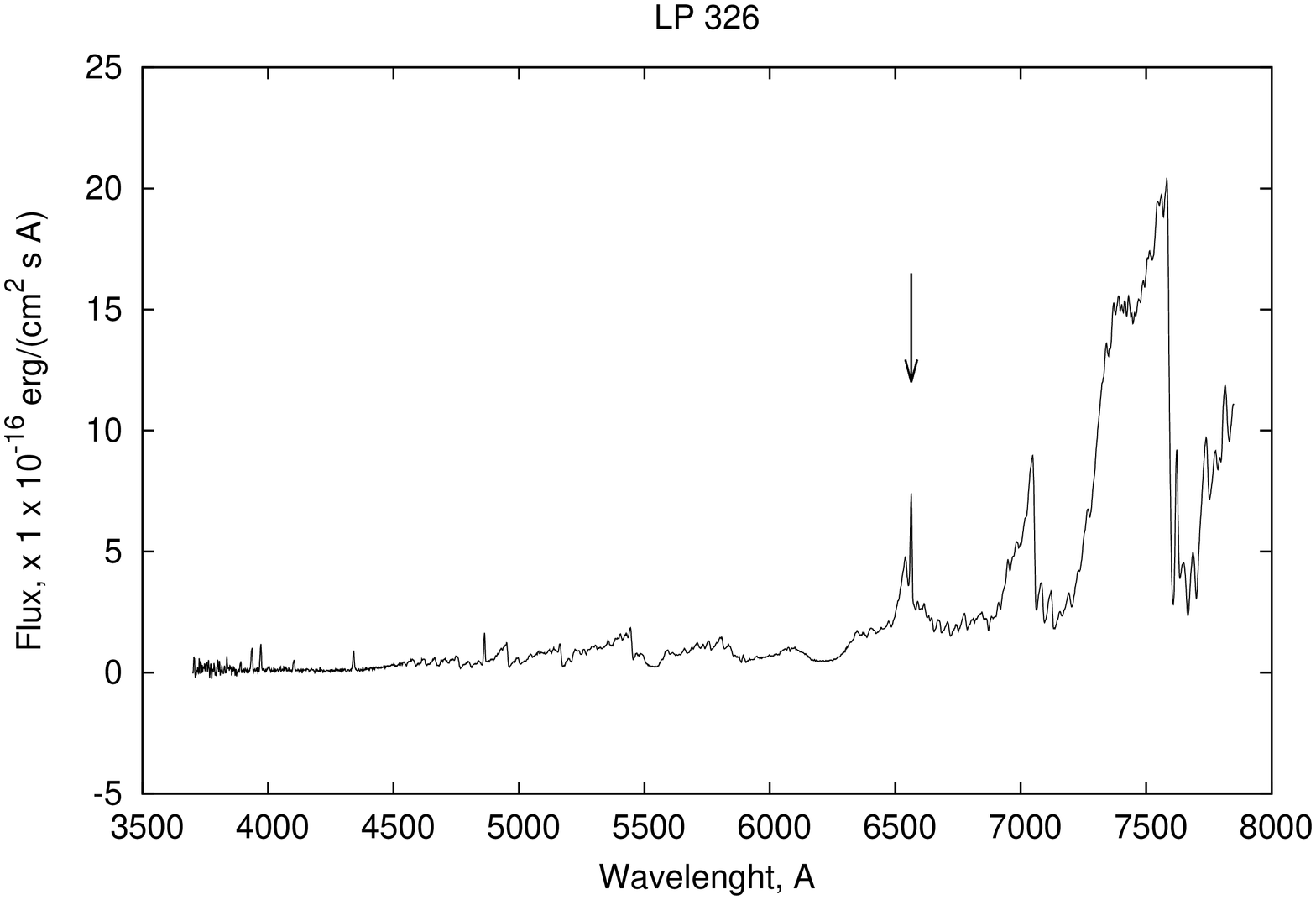} 
\end{array}$ \\

$
\begin{array}{cc}
\includegraphics[width=5.5cm, angle=-90]{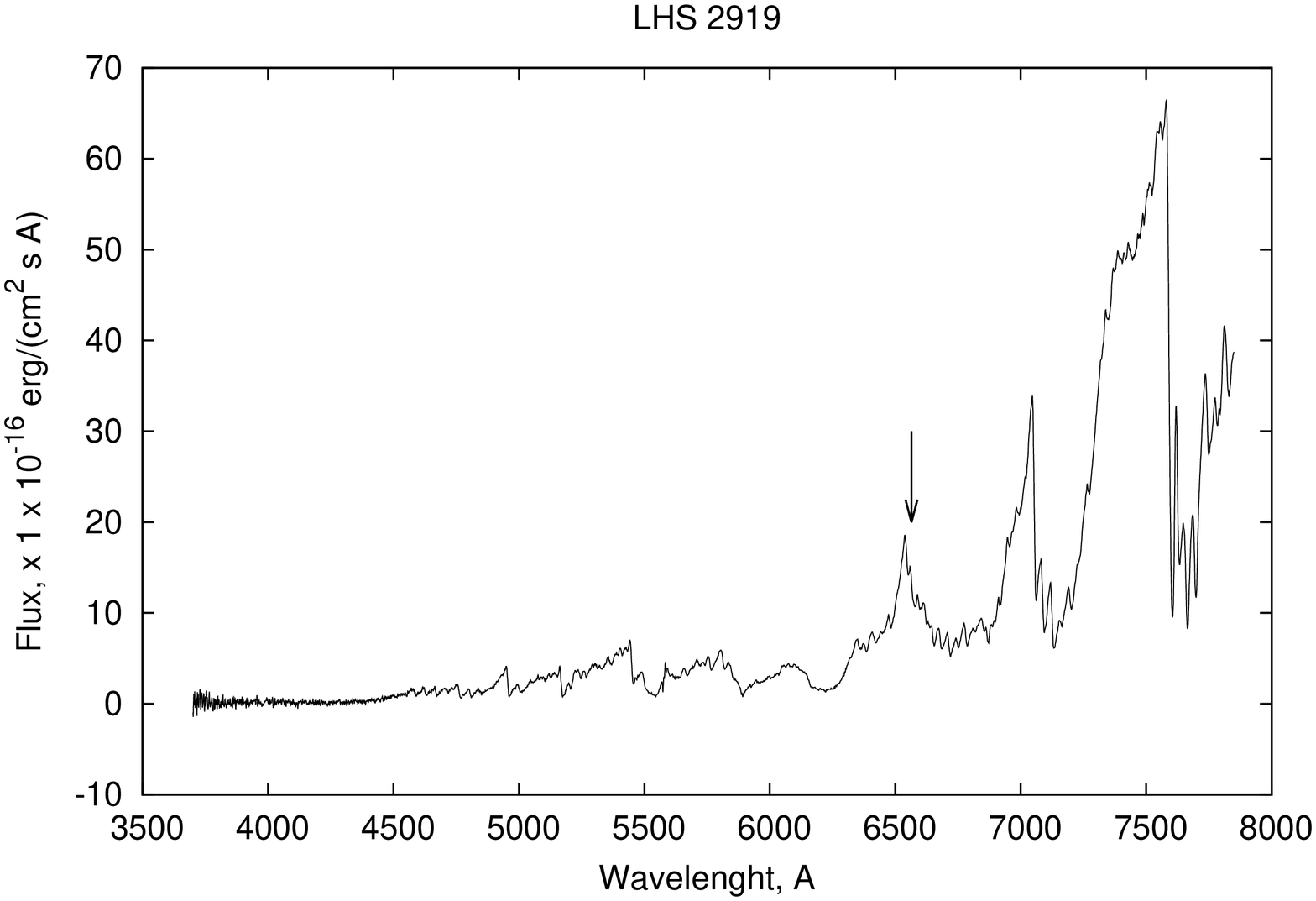} &
\includegraphics[width=5.5cm, angle=-90]{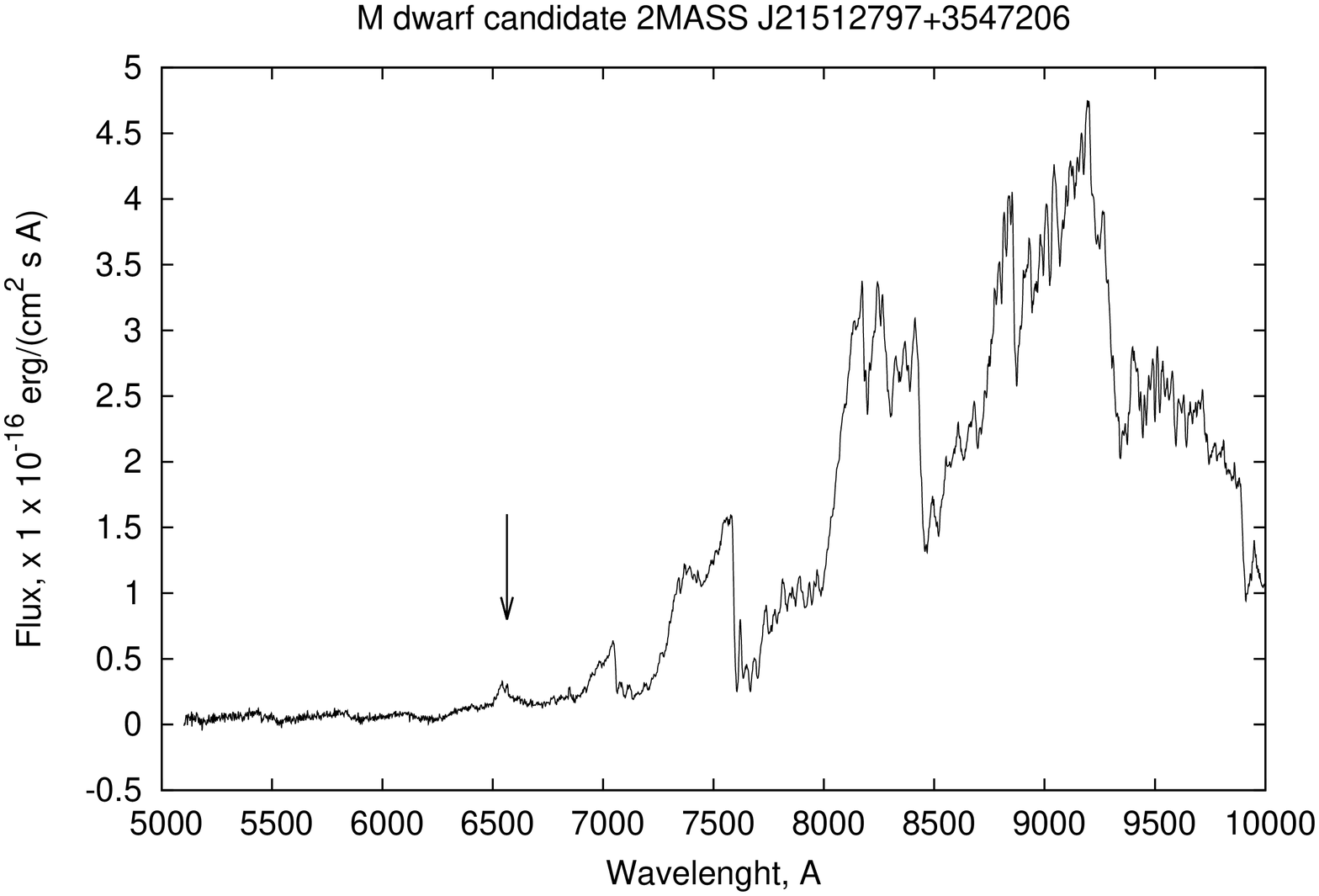} 
\end{array}$\
\label{ha-yes} 
\caption[width=\textwidth]{Top left: spectrum of LP 222-65 where H$\alpha$ emission line is clearly visible. The rest of the Balmer lines are also present, but weaker. Bottom left: spectrum of LHS 2919 with no H$\alpha$ emission. Top right: the spectrum of LP 326-21 with H$\alpha$ in emission. Bottom right: spectrum for 2MASS J20015863+6427486 with weak H$\alpha$ emission.}
\end{figure*}

From the 17 observed dwarfs, 11 show H$\alpha$ in emission. For those 11, we calculated the EW(H$\alpha$) and for the other 6 -- an upper limit. The results are shown in Table~\ref{results}. For comparison the EW(H$\alpha$) calculated by \textsc{hammer} are also presented. For a large fraction of the active dwarfs the rest of the Balmer series are detected, as well as Ca H and K lines, but they have very low S/N. Sample spectra for four objects is given in Fig. \ref{ha-yes}. For those objects with no detectable H$\alpha$ emission, the spectrum around the Ca H and K region had a very poor S/N, thus we are unable to say whether these objects are inactive or have filled-in H$\alpha$ due to low-level activity.\\

\begin{table*}
\begin{minipage}{180mm}
\caption[]{Spectral types, H$\alpha$ equivalent widths and L$_{H\alpha}$/L$_{bol}$ for the studied objects. The M dwarf candidates, marked with * are only partially detected. $\star$ - this object is observed in more than one epoch.
}
\centering
\vspace*{0.3cm}
\label{results}
\begin{tabular}{l c r r r r}
\hline
\hline

2MASS designation    &	Sp.Type	&	EW(H$\alpha$)	&EW err	&	 \textsc{hammer} EW		& $\lg$(L$_{H\alpha}$/L$_{bol}$)	\\
	     	     &	{\tiny /this work/}&	\AA 		& \AA   &	 \AA	  		&	  	\\
			
\hline
       
J13540876+0846083	&	M7.5	&	7.71	&	1.41	&	5.67	&	-4.13	\\
J14280419+1356137	&	M7.5	&	$\lesssim$1.58	&	-	&	$\lesssim$1.49	&	$\lesssim$-4.81	\\
J14441717+3002145	&	M8.5	&	11.89	&	1.58	&	9.50	&	-4.243	\\
J15164073+3910486	&	M6.5	&	10.71	&	1.05	&	8.01	&	-3.97	\\
J16063390+4054216	&	M6.5	&	$\lesssim$0.61	&	-	&	$\lesssim$1.34	&	$\lesssim$-5.21	\\
J16241436+0029158	&	...	&	$\lesssim$5.28	&	-	&	$\lesssim$5.47	&	--	\\
J16342164+5710082	&	...	&	$\lesssim$0.29	&	-	&	$\lesssim$0.15	&	--	\\
J16463154+3434554	&	M6.5	&	$\lesssim$1.04	&	-	&	$\lesssim$1.36	&	$\lesssim$-4.98	\\
J17071830+6439331$\star$ &	M8.5	&	30.36	&	3.13	&	26.94	&	-3.84	\\
				        &	 	&	13.59	&	2.38	&	10.33	&	-4.18	\\
	         			&	 	&	21.62	&	1.66	&	20.61	&	-3.98	\\
J17462953+2903386	&	M7.5	&	14.54	&	4.96	&	10.79	&	-3.85	\\
J17465934+4502473	&	M7.5	&	30.36	&	3.13	&	26.94	&	-3.53	\\
J17502484-0016151	&	L5.0	&	$\lesssim$3.63	&	-	&	$\lesssim$0.66	&	--	\\
J18573008+5015011	&	M7.5	&	11.40	&	2.43	&	8.24	&	-3.95	\\
J20015863+6427486*	&	M7.5	&	4.75	&	2.33	&	4.27	&	-4.33	\\
J21512797+3547206*	&	M9.0	&	2.89  	&	1.76	&	3.17	&	-4.95	\\
J22594403+8013189*	&	M8.5	&	6.69	        &	2.81	&	3.61	&	-4.49	\\
J23333174+7456179	&	M8.0	&	5.43         &	1.20	&	5.16	&	-4.48	\\
 		  
\hline
\end{tabular} 
\end{minipage}
\end{table*}

The flux in H$\alpha$, F$_{H\alpha}$ is also calculated. Next, we calculated the L$_{H\alpha}$/L$_{bol}$ ratio using the $\chi$ method, described in \citet{walkow04}. The parameter $\chi$ is calculated based on hundreds of spectra, and used to calculate the L$_{H\alpha}$/L$_{bol}$ without depending on the distance. For comparison, the objects with known distance $r$ (Table \ref{objects}), the same ratio is calculated using the flux in H$\alpha$. The results obtained using the two methods are consistent and are within the error margin. For example the results for the active dwarf 2MASS J17071830+6439331 are\\

\noindent with the $\chi$ method: L$_{H\alpha}$/L$_{bol}$ = (1.46 $\pm$ 0.15) $\times$ 10$^{-4}$, \

\noindent with $F$(H$\alpha$)~ and~$r$: L$_{H\alpha}$/L$_{bol}$ = (1.61 $\pm$ 0.17) $\times$ 10$^{-4}$. \\

\begin{figure*}
  \begin{minipage}{180mm}
\begin{center}
\epsfig{file=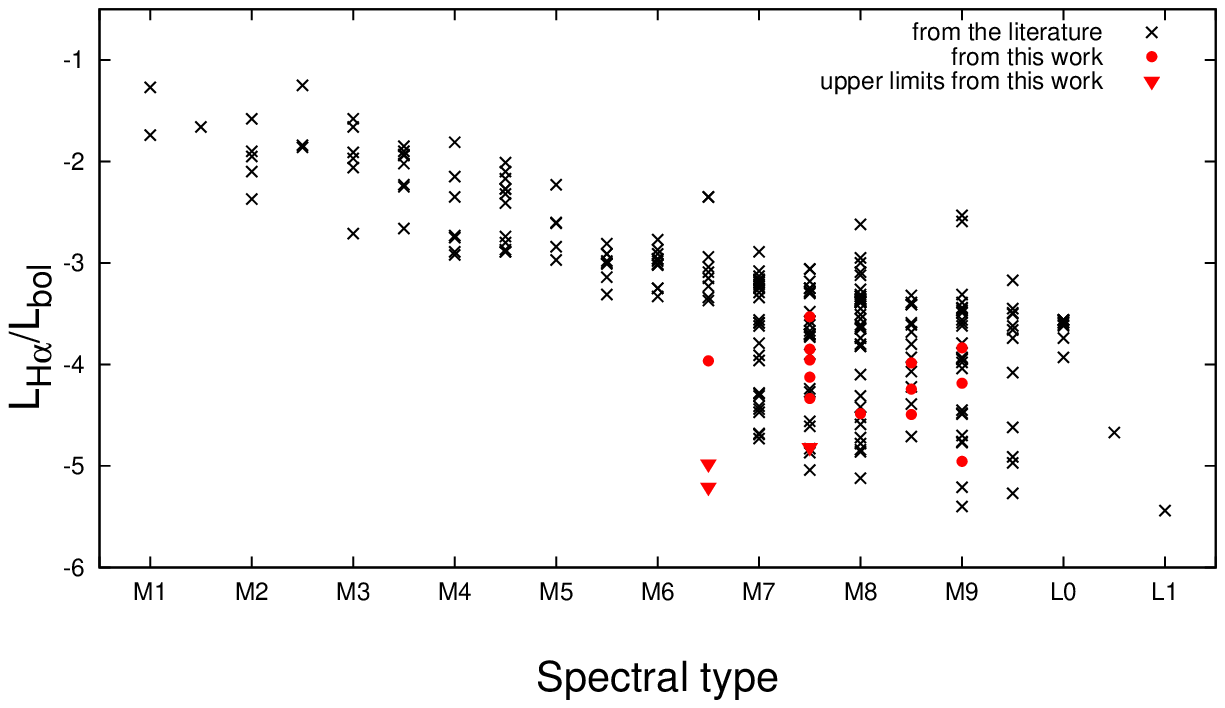, width=14cm}
\vspace{-0.5cm}
\caption[]{log(L$_{H\alpha}$/L$_{bol}$) -- spectral type relation. Black symbols represent the data from the literature \citep{gizis00,mclean12}, red dots -- the measurements from this work and red triangles -- the calculated upper limits.\
}
\label{spt-Ha} 
\end{center}
\end{minipage}
\end{figure*}

In Table \ref{results}, we systemize the determined spectral types and L$_{H\alpha}$/L$_{bol}$ ratios. Fig. \ref{spt-Ha} represents sample of objects from the literature (black) and the results of this work (red). The results obtained in this work correspond well with data from the literature. The determined upper limits are lower than the corresponding values for the given subclass. For three of the objects where H$\alpha$ is present, but with a weak intensity (2MASS J20015863+6427486, 2MASS J21512797+3547206 and 2MASS J22594403+8013189,  marked with * in Table\ref{results}), further observations are required.\

\subsection{Objects of interest}

Here we discuss two interesting objects - LP 326-21 (2MASS J14441717+3002145) and 2MASS J17071830+6439331.
For both dwarfs we found EW(H$\alpha$) measurements in the literature from previous epochs.\

\begin{itemize}

\item LP 326-21 is an M8.5 dwarf with EW(H$\alpha$) = 7.4 \AA~ from \citet{gizis00}. Compared with our observation from 2012, EW(H$\alpha$) = 11.9 $\pm$ 1.6 \AA\ (Fig. \ref{ha-yes}), is noticeable that the dwarf remains with relatively constant equivalent width (within the error), i.e. it possibly maintains quasi constant activity over long periods of time (decades). 

\item 2MASS J17071830+6439331 is M8.5 dwarf with H$\alpha$ flares observed in previous epochs and also in this work. Those flares increase the luminosity in H$\alpha$ by orders of magnitude and have durations in order of minutes.
The spectral observations in the literature are from several epochs - in 1999 with measured EW(H$\alpha$) = 9.8 \AA\ \citep{gizis00}, and in 2003 with measured EW(H$\alpha$) = 28.7 \AA\ \citep{schmidt07}, probably during a flare. 
\end{itemize}

\begin{figure*}
 \begin{minipage}{180mm}
\begin{center}
\epsfig{file=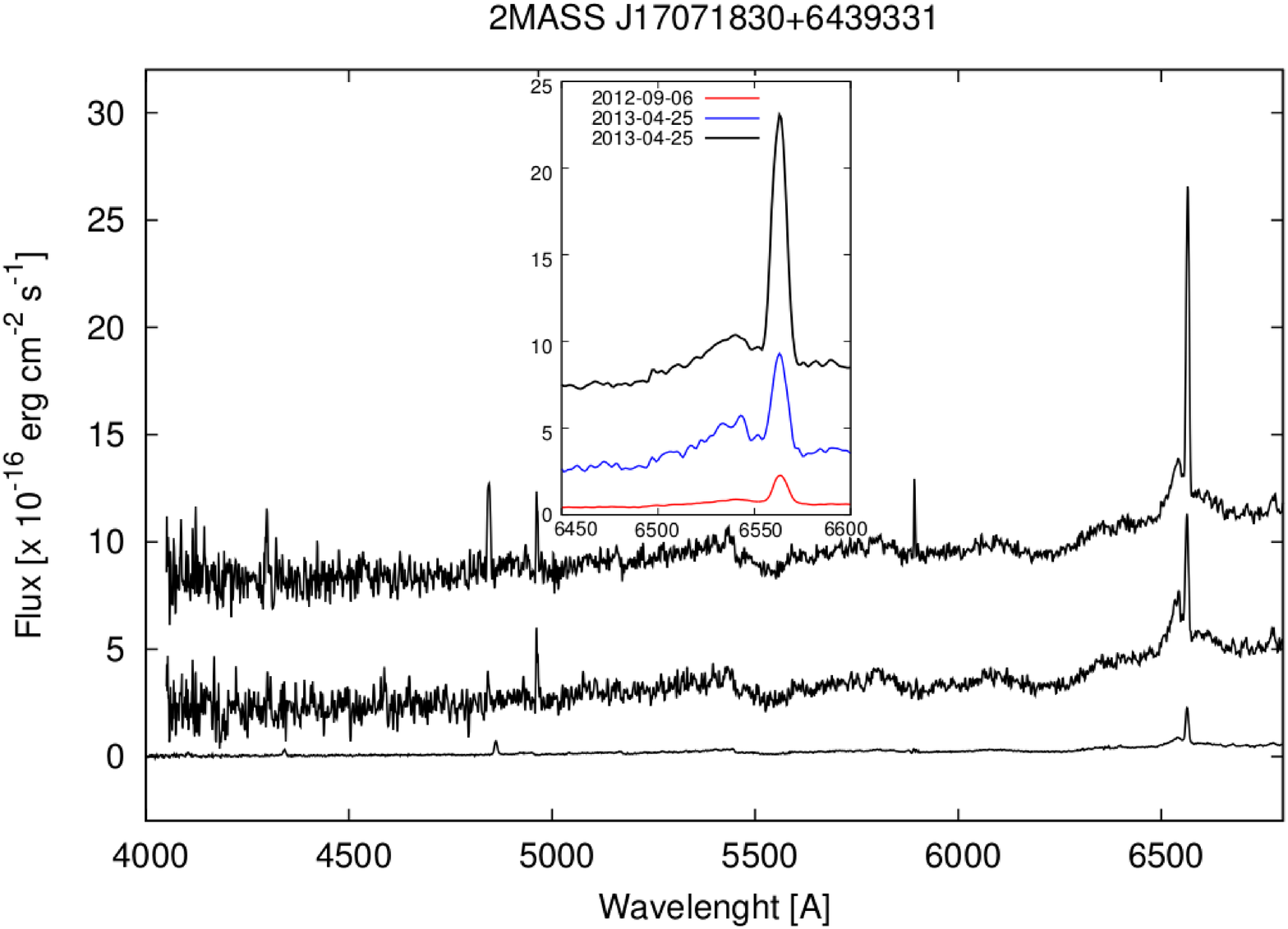, width=14cm}
\caption[]{The variation in the whole spectrum of 2MASS J17071830+6439331, and around the H$\alpha$ emission line (inset).\
}
\label{4000-6750} 
\end{center}
\end{minipage}
\end{figure*}

Similar behaviour was seen in this work where data taken in 2012 September had EW(H$\alpha$) of 21.6 \AA. Observations in 2013 April, showed a raise in the equivalent width up to 30.4 \AA\ in the first frame, followed by a drop to 13.6 \AA\ in the second frame, obtained 10 min after the beginning of the first one; this was probably the end of a flare. Fig. \ref{4000-6750} presents the variation around the H$\alpha$ line and in the whole spectrum of the dwarf. In addition, \citet{rockenf06} report photometric variability with a period of 3.6 h, which they attribute to modulation of the light curve due to magnetic spot rotation. Since the period above agrees well with the measurements for \textit{v}$\sin$\textit{i}, this may be the period of rotation of the dwarf. During the same observations the authors also registered a UV burst, which is characteristic of enhanced magnetic activity. \

\section{Discussion}

The study of magnetic activity is essential for understanding of the physical properties and characteristics of the objects around and below the stellar/substellar boundary. To fully understand the processes occurring in these objects and the role of activity in the evolution, it is important to conduct a number of observations and studies of more physical parameters. In this work, original observations of 17 late-type objects are presented. Seven of those are spectrally observed for the first time and classified as late-M dwarfs. For 15 of the objects the EW(H$\alpha$) and L$_{H\alpha}$/L$_{bol}$ (or upper limits), characterizing the activity of the dwarfs, are calculated. These results are an important addition to the available literature information and allow more complete and accurate characterization of the magnetic activity in cool dwarfs. For two of the observed dwarfs, data from previous epochs are available, and are now enriched with new observations. Both of those dwarfs show the presence of magnetic activity for long periods -- in the first case quasi-constant-like and in the second -- variable. The zero H$\alpha$ dwarfs could mean either a very weak/zero chromosphere or that the chromosphere is of moderate strength with the H$\alpha$ line filled-in. \citet{MathioudakisDoyle} showed that Mg~{\sc ii} 
H and K is an excellent line to provide this additional diagnostic, unfortunately such an ultraviolet instrument does not currently exist. \

\section{Acknowledgments}

We gratefully acknowledge funding for this project by the Bulgarian National Science Fund (contract no. DDVU02/40/2010). This research is based on 
observations made with the GTC, installed in the Spanish Observatorio del Roque de los Muchachos of the Instituto de Astrof\'{\i}sica de Canarias, in the island of La Palma. We use data from the SIMBAD data base, operated at CDS, Strasbourg, France, plus data products from the Two Micron All Sky Survey, which is a joint project of the University of Massachusetts and the Infrared Processing and Analysis Center/California Institute of Technology, funded by the National Aeronautics and Space Administration and the National Science Foundation.  The Digitized Sky Surveys were produced at the Space Telescope Science Institute under US Government grant NAG W-2166. The images of these surveys are based on photographic data obtained using the Oschin Schmidt Telescope on Palomar Mountain and the UK Schmidt Telescope. The plates were processed into the present compressed digital form with the permission of these institutions. The National Geographic Society - Palomar Observatory Sky Atlas (POSS-I) was made by the California Institute of Technology with grants from the National Geographic Society. This research has benefitted from the M, L, T and Y dwarf compendium housed at DwarfArchives.org. Research at the Armagh Observatory is grant-aided by the Northern Ireland Departament of Culture, Arts and Leisure.\


\bsp

\bibliographystyle{mn2e}

\bibliography{gtc_v6}

\label{lastpage}

\end{document}